\documentclass[aps,pre,preprint,groupedaddress,showpacs]{revtex4}

\usepackage{graphicx}
\begin{document}


\title{A Novel Approach to Complex Problems}



\author{Shmuel Nussinov}
\email[]{nussinov@ccsg.tau.ac.il}
 \affiliation{
Department of Physics\\
Johns Hopkins University \\
Baltimore MD 21218\\
and\\
Tel-Aviv University, School of Physics and Astronomy \\ 
Tel-Aviv, Israel \\
}
\author{Zohar Nussinov}
\email[]{zohar@lorentz.leidenuniv.nl}
\affiliation{Institute Lorentz for Theoretical Physics, Leiden University\\
POB 9506, 2300 RA Leiden, The Netherlands\\
}


\date{\today} 

\begin{abstract}
A novel approach to complex problems has been previously applied to
graph classification and the graph equivalence problem.  Here we
consider its applications to a wide set of NP complete problems,
namely, those of finding a subgraph $g$ inside a graph $G$.
\end{abstract}




\maketitle



\section{Introduction}

In many areas of physics, biology, economics and other fields we
encounter ``Complex Problems". Such problems, like finding ground
states in spin glasses\cite{Meza} (or ``frustrated" systems in
general), folding proteins once the sequence of amino acids is known,
or ``The Traveling Salesman" problem\cite{Textbook} (and the closely
related ``Hamiltonian Path" problem for graphs) may appear to be very
different. Yet the complexity of these and a host of other problems
stems from a common source: There are many, often conflicting,
requirements on the elements that embody the solution.

An archetypical\cite{Textbook} example is the ``Satisfiability
Problem",\cite{Goldstone} where $O(n)$ requirements are 
imposed on $n$ elements.
We are asked if all can be simultaneously satisfied, and
if so, to actually exhibit the solution.
The difficulty in trying to piece out a solution is that we need
to make discrete, binary (yes or no) decisions at each step.
We are guided by the underlying constraints and/or the
desire to minimize the overall ``energy"or ``cost". Yet due to the
frustrations/conflicts there is no clear strategy for
making such choices so as to consistently move towards the solution.
``Improvements" obtained at some stage may be later undone and
particular choices may exclude other choices many steps later.

Thus if certain pairs of amino acids are near each other in a
tentative folded protein, then the demands imposed by the primary
sequence and/or excluded volume prevent us from bringing certain other
pairs into close proximity.
Also visiting a certain vertex excludes, by the very
definition of the Hamiltonian path, its future use
in trying to complete our tour of the graph.

We suggest a novel approach for addressing such problems:
\begin{itemize}
\item[(i)] Instead of performing discrete, maximal, changes of just two
elements at a time, we perform at each stage small changes of all
elements.

\item[(ii)] The original system is mapped onto a simple ``physical"
model with $n$ degrees of freedom. To have maximal symmetry so as to
avoid  biasing in the initial state, and in order to 
minimize/exclude frustrations, the model is embedded in 
$d \approx n$ dimensions.

\item[(iii)]We endow our model with a first-order, deterministic dynamics
and (numerically) evolve it for some time. The ``dynamics", i.e, the
forces between the various elements, is chosen in such a way that
if a solution exists, then the initial model evolves into a specific
final form from which the solution is readily inferred.
\end{itemize}
\noindent Ideally we will be able to show that this can be achieved in a
polynomial number of steps.

The above are general principles. For each problem we
need to specify the model, its dynamics and prove/verify that it
indeed achieves the solution (hopefully in polynomial time).
To date the above approach has been applied to two problems in 
graph/network theory:

First [GJN]\cite{GJN} used a symmetric $n$ simplex model in $n$ dimensions. 
Its evolution identifies in polynomial number of steps via geometrical
bunching of sets of points, various ``clusters" or ``Imperfect Cliques"
in the graph and  assesses the ``Communication Distance" between them
(Clusters are groups of vertices with higher than average mutual
connectivity.)

Second [GN]\cite{GN} have shown that the same algorithm solves, in even a
shorter time, the graph equivalence problem (GEP) which is defined as
follows:  A non-directed graph $G$ with $n$ vertices is specified by a
connectivity matrix $C(i,j)$ with entries 1 (or 0) if
vertices $i$ and $j$ are (not) connected.
We are given two such connectivity matrices  $C$ and $C'$.
Are these matrices representing  the same graph and, in the case
that they do, can we exhibit the permutation 
which makes them identical?

The first problem of approximate graph characterization (AGC) has
considerable practical importance as many networks are likely to have
``Hidden Clusters".  Also the GEP was not solved to date in polynomial
time. 

Can our novel approach be applied to the NP complete
problems?\cite{Textbook} The difficulty of these problems could
manifest---albeit in rather subtle ways which are of interest in their
own right---also in the present approach, preventing a solution in a
polynomial time.  Thus for {\it any} physical model and {\it any}
dynamics that we invent in our effort to solve an NP complete problem,
an extremely long time, say, $t=exp(a.n)$, could be required in order
to converge to the solution.

This would be the case if dynamical models emulating any NP complete
problem necessarily have chaotic unstable motion. Alternatively the
true minimum energy state of the system--which is the desired
solution--could be masked by a plethora of local minima in which our
system will be trapped almost indefinitely.  We cannot exclude the
above possibilities.

Still we conjecture that our novel approach is applicable to NPC
problems.  We are presently attempting to prove this by actual
implementation of this new approach in specific codes.

All the problems of finding a specific sub-graph $g$ (with $n$
vertices) within a graph $G$ (with $N$ vertices) are NPC.
These include finding the largest perfect clique and finding a
Hamiltonian circuit.
These problems can be reformulated in our approach as searching
a perfect dynamically generated ``Docking" of a physical model of
$g$ onto some subset of vertices/edges of $G$. This is discussed in
Sec. III, utilizing evolution in $(N+n)$ dimensions.

\section{Brief Review of Previous Applications}

To introduce notation and illustrate the ideas, we briefly review
earlier applications to the AGC\cite{GJN} and GEP\cite{GN}
problems.
Consider a symmetric $n$ simplex in $(n-1)$ dimensions whose
vertices $\vec{r}(i)$, represent the abstract vertices $V(i)$ of the 
graph.\cite{6}
We introduce attractive (repulsive) interactions between the
points $\vec{r}(i)$ and $\vec{r}(j)$ if the corresponding vertices are
(or are not) connected in the original graph:
\begin{equation}
U = \sum_{i>j} U_a (| \vec r_i - \vec r_j |)C_{ij}+ \sum_{i>j} 
U_r (|\vec r_i - \vec r_j|)(1-C_{ij})
\end{equation}  
The $n$ points then move under the resulting forces:
\begin{equation}
\mu \frac{d\vec r_i(t)}{dt}=\vec F_i (\vec r_i (t))
\end{equation} 
according to
\begin{equation}
\vec F_i = -\vec \nabla _{(\vec r_i)} \{U \left[ \vec r_i, \cdots, \vec r_n  \right] \}
\end{equation}\\ 
The first order dynamics ensures that the system
consistently moves in the fastest way towards its minimum without
the ``overshoots" and oscillations of the second-order formulation.
In $(n-1)$ dimensions all the distances $|\vec{r}(i)-\vec{r}(j)|$ are 
independent (apart from constraints due to triangular 
inequalities). A monotonic $u(r)$
then excludes any local maxima, minima or saddle points.
Thus the system cannot be trapped in any local minimum and will attain
its absolute minimum at a boundary point.

After a sufficiently long evolution the original symmetric simplex
will be significantly distorted in a way which reflects the topology
of the graph in question.
Specifically, points $\vec{r}(i)$ representing vertices $V(i)$  which are
``close in the graph", i.e., have many short paths connecting them
in the graph, tend to move closer in space.
Conversely, points which are ``far in the graph", i.e., have only longer
connecting paths tend to move apart.

The original set of $n(n-1)/2$  $C(i,j)$'s then gets mapped
onto the $n(n-1)/2$ distances $|\vec{r}(i)[t]-\vec{r}(j)[t]|= R(i,j)$.  An identical
evolution of an equivalent (permuted)
$C'(i,j)$ yields another ``distance-matrix", $R'(i,j)$. $R'$ is obtained
from $R$ via the same similarity permutation which maps $C$ into $C'$.
In general the elements of $R$ (or $R'$) are---unlike those of
$C$ and $C'$---all different. Verifying that $R$ and $R'$ are
obtained by vertex re-labeling, and finding the required permutation
can then be easily done in $O(n)$ steps.

The dynamical evolution above is simulated by using:
\begin{equation}
\vec r_i (t + \delta ) = \vec r_i (t) + \frac {\delta }{\mu } \vec F_i (\vec r_e (t))
\end{equation} 
Altogether there are $n \cdot v$ edges in the graph with $v$ the
average valency (the valency is the number of edges impinging at a
vertex).  In each simulation step we compute $n \cdot v$ forces
$F(i,j)$. Since each force has $(n-1)$ components we need altogether
$O(n^2 \cdot v)$ computations per step. The ``size" of the problem $n$
is thus (polynomially) reflected in the number of computations in each
step.  Note, however, that since in each iteration step {\it all} the
points move, the number of iteration steps, $s$, required for graph
comparison, and/or clustering, need not grow at all with $n$.  In many
cases the valency $v$ is also $n$ independent and the complexity of
the algorithm is only $O(n^2)$!

\section{Applying the Novel Approach to NP Complete Problems} 

So far we utilized motions of the $N$ points of an $N$ simplex
in $N-1$ dimensions to model problems associated with one graph
$G$ with $N$ vertices. Also to test if $G$ and $G'$ are equivalent 
we separately
processed the corresponding simplexes $S$ and $S'$.

For the task of finding a replica of a ``small" graph $g$ with $n$
vertices inside a ``big" graph $G$, we introduce one extra $(n-1)$ dimensional $n$ simplex
$s$, corresponding to $g$.
To better emulate certain complex problems we may  allow $S$ and
$s$ to be non-symmetric and non-rigid. For now we keep
both $S$ and $s$ rigid, symmetric and with a common edge length:
\begin{equation}
   |\vec{R}(I)-\vec{R}(J)|=a=|\vec{r}(i)-\vec{r}(j)| \;\; 
I \neq J=1...N; i \neq j=1...n
\end{equation}
Let $C(I,J)$ and $c(i,j)$ be the connectivity matrices for $G$ and
$g$, respectively. The small graph $g$ can be found inside $G$ if,and only
if, a combinatorial perfect docking of $s$ inside $S$ is possible. The
latter is defined as follows:

All edges of $g$ represented in s by $\vec{r}(i)-\vec{r}(j)$ with
$c(i,j)=1$, match edges in $G$ represented in $S$ by
$\vec{R}(I)-\vec{R}(J)$ with $C(I,J)=1$.  Likewise missing edges in
$g$ should also correspond to missing edges in $G$.  In our model it
means that the representative vectors in $s$, namely,
$\vec{r}(i)-\vec{r}(j)$ with $c(i,j)=0$ should dock onto
$\vec{R}(I)-\vec{R}(J)$ with $C(I,J)=0$.

Our choice of symmetric, equal edge simplexes $S$ and $s$ allows a
geometricl match of $s$ with any symmetric $n$ sub-simplex; i.e., any
$(n-1)$ dimensional ``Face" of the large simplex $S$.  We next introduce
interactions between the edges of $s$ and those of $S$ which
correspond to true edges in $g$ and $G$. This is done in such a way
that a perfect combinatorial docking of $s$ inside $S$ constitutes the
lowest energy state of the system.  Thus imagine putting at the
centers, $\{\vec{R}(I)+\vec{R}(J)\}/2$, of the edges in $S$ for which
$C(I,J)=1$, positive, unit charges. Likewise, we put at all
$\{\vec{r}(i)+\vec{r}(j)\}/2$ for which $c(i,j)=1$, negative, unit
charges.  The total mutual interaction energy then is:
\begin{equation} 
U\{R(I),r(i)\} = - \sum^N_{I \neq J} \sum^n_{i \neq j} 
C(I,J) \cdot c(i,j) \{V(|\vec{R}(I)+\vec{R}(J)-\vec{r}(i)
-\vec{r}(j)/2|\} 
\end{equation}
For any pairwise potential $V(\rho)$ which is monotonically
decreasing with $\rho$ the absolute
minimum of the interaction energy occures when $s$ has perfectly docked
inside $S$. This is when $(\vec{R}(I)+\vec{R}(J)-\vec{r}(i)-\vec{r}(j))$ vanishes for all
\begin{equation}
                          C(I,J) \cdot c(i,j)=1.
\end{equation}
Keeping $S$ fixed at some standard position we let $s$ move according
the first-order dynamics  (Eq. (4)) under the attractive, pairwise
forces which derive from the above potentials.
Since $s$ is rigid, this motion consists mainly of rotations.
This requires some relatively straightforward modifications
with (anti-symmetric) tensor torques $\tau(\alpha,\beta)$ and
corresponding infinitesimal  rotations $\theta(\alpha,\beta)$ in the
$\alpha,\beta$ plane replacing vector forces and  
displacements\cite{7}:
\begin{equation}
\theta _{\alpha \beta }(t + \delta ) = \theta _{\alpha \beta }(t) + 
\frac {\tau _{\alpha \beta }}{\mu } \delta
\end{equation}  

Imagine that we are performing all of this in $(N-1) > (n-1)$ dimensions.
To avoid any biasing we should place $s$, at $t=0$, in a
position which is completely symmetric with respect to $S$.
Thus whatever non-symmetric movements of $s$ eventually occur these
will solely reflect--via the asymmetric forces/torques exerted on
$s$--the asymmetric $C(I,J)$.

Unfortunately we have $(\stackrel{^N} ~\!\!\!\!\!_{n})$
different ways of putting $s$ at the center of $S$
and orienting it parallel to any one of the $(n-1)$ dimensional ``Faces"
of $S$ which are symmetric $n$ (sub)simplexes each.

Since every such choice may bias the outcome of the dynamical
evolution utilizing these particular initial conditions, we should
repeat the process for each of the ($\stackrel{^N}~\!\!\!\!\!\!_{n}$)
different orientations, thereby clearly exhibiting a non-polynomial
complexity.  

This difficulty is avoided by formulating the problem 
in $(N+n-2)$ dimensions, the sum
of the dimensions $(N-1)$ and $(n-1)$ of the embedding spaces of $S$
and $s$.

From the vantage point of the $(N-1)$ dimensional ``Big" simplex $S$,
we are putting all of the simplex $s$ at the only symmetric location,
namely just at its center!  We choose the latter to be the origin of
the $(N-1)$ cartesian coordinate system: $X(1)$, $X(2)$,...,$X(N-1)$.
Now it is immaterial how we choose to orient $s$ relative to the
remaining $(n-1)$ coordinates: $x(1)$, $x(2)$,...,$x(n-1)$ whose
origin we again fix at the same common location.  The point is that
all the forces depend only on distances $|\vec{R}(I)-\vec{r}(i)|$.  By
the Pytagorean theorem these depend initially only on sums of squares
of $x(i)$ and $X(I)$ coordinates and are therefore invariant with
respect to any rotation of the $x(1)$,...,$x(n-1)$ coordinate frame.

Once the torques due to the forces between pairs of edges in $S$ and
$s$ start operating this initial symmetry will be broken and the
system will evolve (mainly rotate) in the full $(N+n-2)$ dimensions.
The energy is then a function of a large $\{(N+n-2)\cdot(N+n-3)/2\}$
number of rotation angles.  We conjecture, but have not been able to
yet prove, that, as in the case of translations, this large number of
degrees of freedom avoids local minima of the energy.  This would
imply that the system persistently eveolve in short time towards its
absolute minimum.  The latter is achieved on the ``boundary" namely
when all the corresponding vertices in $S$ and in $s$ overlap,  
and $g$ has been found in $G$.

Many NP complete problems correspond to special choices of the ``small
graph $g$.  Having all $c_{ij} = i \neq j = 1...n$ corresponds to the
problem of searching for a perfect clique of size $n$ inside
$G$.\footnote{It appears (V. Gudkov and S.Nussinov, unpublished) that
for this specific problem, even the "old" method of processing only
the {\it one} original graph $G$ may be applicable.  We ``tune" the ratio of
strengths of the repulsive and attractive interactions.  By using a
ratio of order unity---which is considerably higher than the one
previously used in Ref. \cite{GJN} where we looked for clusters or
``imperfect cliques"---we selectively collapse only (and hopefully
all) the perfect cliques in the graph $G$.}
Taking $(n=N)$ and $c_{12} = c_{23} = ... c_{j,j+1} = ...  c_{n-1,n} 
= c_{n,1} = 1$, all other $c_{ij} = 0$---i.e., having $g$ which consists
of just one cycle of $(n=N)$ vertices---yield the Hamiltonian path
problem; namely, of searching a Hamiltonian path in 
$G$.\footnote{In this specific case, since we associate with 
both $G$ and $g$  symmetric $(n=N)$  simplexes $S$ and $s$, 
we could put the simplex $s$ so as to literally overlap $S$ 
coexisting in the same $(N-1)$ dimensional space.
The initially chosen orientation (or the labeling of vertices in 
$g$ used
in the above defintion of the connectivity matrix of $g$) will,
in general, {\it not} correspond to the minimal energy.
The simplex $s$ will then start rotating due to the forces
constructed above relative to $G$ until the ``combinatorially
(= energetically) optimal" docking obtains.
Since only the even permutations of $N$ elements can be achieved via
continous rotation we may need to repeat our search twice 
starting once
with the simplex $s$ and once with $s'$; namely, $s$ in which {\it one}
transposition, say, $V(1) \leftrightarrow V(2)$ has been performed.}
The latter
is a path in $G$ consisting of edges $E_{I,J}$ in $G$, which visits
{\it all} the vertices in the graph exactly one time.  This is the
simplest variant of the celebrated ``Traveling Salesman problem" where
all the distances between any pair $g$ cities are the same,
highlighting the purely combinatorial (rather than algebraic aspects)
of this NPC problem.

\section{Summary}

We briefly describe a new deterministic, computational approach
to complex problems and conjecture 
that the new approach may be applicable to
NPC problems.  In particular, we 
considered the problem of searching a graph $g$
inside $G$. Jointly with V. Gudkov we are presently testing the idea
in concrete codes to see what is the real impact on some NPC problems.  If successful, our approach will have many applications in quite a few 
areas.\footnote{It has been argued that by studying neural nets and the
human brain in particular, we may artificially emulate the vesatility,
pattern recognition and problem solving powers of the latter. We would
like to suggest that the working of our brain is similar to that of
our novel approach, though there is no analogue physical system there
tuned to solve common problems.  The highly inter-connected system of
neurons has---unlike ordinary CPU's and even more advanced special
purpose ``lattice" computers designed for specific physical problems
(eg. nonpertubative QCD)---a high effective (Hausdorf) dimension. Also
our distributed memory and the intuitive sensing of a complete
``Gestalt" form strongly suggests a mode of operation with almost
continous, small, changes of many elements in parallel.}

\begin{acknowledgments}
S.N. would like to dedicate this work to Sir I. Wolfson who donated
the chair in theoretical physics at Tel-Aviv University on the
occasion of his 80th birthday.
\end{acknowledgments}



\begin{thebibliography}{99}
\bibitem{7} Strictly the analog of the first-order dynamics for 
translation (Eq. (2)) is 
$\mu_{\alpha\beta,\sigma\nu}\omega_{\sigma\nu} = \tau_{\alpha\beta}$
with $\tau _{\alpha \beta } = \sum_{ij, i'j'} (F_{iji'j'})_{\alpha } (\rho _{ij})_{\beta }-(\beta \leftrightarrow \alpha )$
the torque exerted by $S$ on $s$. The viscosity tensor $\mu _{\alpha \beta \sigma \tau } $ reduces
to a constant if we assume that the frictional forces act symmetrically at all the $n$ vertices of $s$.  The last equations then becomes
$\tau _{\alpha \beta } = \mu \omega _{\alpha \beta}$.
\end{thebibliography}
\end{document}